\documentclass[aps,prb,showpacs,reprint,superscriptaddress]{revtex4-1}

\usepackage{graphicx}
\usepackage{color}
\usepackage{amsmath}
\usepackage{bbm}
\usepackage{amssymb}

\usepackage{dsfont}

\usepackage[utf8]{inputenc}	
\usepackage[T1]{fontenc}

\input epsf

\begin{document}

\title{Interplay between the extended s-wave symmetry of the gap and the spin-orbit coupling in the low-electron concentration regime of quasi-two-dimensional superconductors}
\author{M. Zegrodnik}
\email{michal.zegrodnik@agh.edu.pl}
\affiliation{Academic Centre for Materials and Nanotechnology, AGH University of Science and Technology, Al. Mickiewicza 30, 30-059 Krakow,
Poland}
\author{P. W\'ojcik}
\email{pawel.wojcik@fis.agh.edu.pl}
\affiliation{AGH University of Science and Technology, Faculty of Physics and Applied Computer Science,30-059 Krakow, Poland Al.  Mickiewicza 30, 30-059 Krakow, Poland}


\begin{abstract}
We analyze the real-space paired state with the $\mathbf{k}$-dependent superconducting gap in the presence of Rashba type spin-orbit coupling and external magnetic field. We show that the $extended$ $s$-$wave$ pairing symmetry is the most probable scenario to appear in the low-electron concentration regime. According to our study, the van Hove singularity induced by the spin-orbit coupling may lead to a significant enhancement of the superconducting gap, critical temperature and critical magnetic field. Moreover, the combined effect of the spin-orbit coupling and the external magnetic field results in a non-zero total momentum of the Cooper pairs, which is a characteristic feature of the so-called helical state. In such situation, due to the C$_4$ symmetry breaking, a small $d$-$wave$ and $p$-$wave$ contributions to the pairing appear, which significantly change the character of the helical state. The obtained results are discussed in the context of the experimental data related with the unconventional superconducting features of the transition metal oxide interfaces as well as the recently reported supercurrent diode effect.
\end{abstract}

\maketitle

\section{Introduction}
In recent years the quasi-two dimensional superconducting systems have attracted growing interest due to their rich physics as well as high degree of tunability\cite{Uchihashi_2016}. In particular, the two-dimensional superconductivity has been reported in monoatomic films\cite{Zhong2016,Fatemi2018,Zhang2020} as well in both LaAlO$_3$/SrTiO$_3$ and LaAlO$_3$/KTaO$_3$\cite{Chen2021,Liu2021,Citro_2021} interfaces.

In connection to many of the mentioned systems the $extended$ $s$-$wave$ pairing symmetry has been discussed. In particular, as shown by us recently such symmetry of the superconducting gap may cause the characteristic dome-like shape of T$_C$ as a function of gate voltage in the LaAlO$_3$/SrTiO$_3$ interface in good agreement with the experimental data\cite{Zegrodnik2020,Wojcik2021}. Moreover, it has been shown theoretically that in the regime of low electronic concentrations realized in CuO monolayers the $extended$ $s$-$wave$ symmetric gap appears with a similar dome-like behavior as a function of hole doping\cite{Jiang2018,Zegrodnik2021}. The same gap symmetry leading to the so-called $s^{\pm}$ paired state is also discussed in the context of monoatomic FeSe\cite{Zhang2020} or the superconducting FeAs layers of the popular iron-pnictides. 

Among the aforementioned group of quasi two-dimensional superconducting systems, particularly interesting are the LaAlO$_3$/SrTiO$_3$ and LaAlO$_3$/KTaO$_3$ interfaces which are characterized by an electrically tunable $T_C$ as well as significant spin-orbit coupling (SOC)\cite{Reyren2007,Rout2017, joshua2012universal,maniv2015strong,Chen2021,Liu2021}. The latter leads to an increase of the critical magnetic field well above the paramagnetic limit\cite{Wojcik2021,Rout2017}. At the same time, it has been argued that in the two dimensional systems in the presence of the in-plane magnetic field, the non-zero momentum Cooper pairing may appear leading to the widely known Fulde-Ferrell-Larkin-Ovchinnikov state (FFLO)\cite{Zheng2014,Wojcik2019}, which is strengthen by the spin-orbit coupling and than takes the form of the so-called helical state \cite{Matsuda2007}.
In the past decades great efforts have been made to detect signatures related with the FFLO phase. However, up to this day the direct experimental evidence is lacking. One important reason for that is the fact that most probably the FFLO state exists in a narrow parameter regime near the critical point. Therefore, it may be simply missed out by the experimental analysis. Also, the disorder effects in superconductors induce strong scattering between different momenta that destroys the superconducting pairing. However, in the transition metal oxide interfaces the entire phase diagram can be easily scanned by tuning the gate voltage thus avoiding the disruptive effects of chemical doping. Therefore, those systems seem to be good candidates for both the appearance and detection of the FFLO state.

The research focused on the interplay between $extended$ $s$-$wave$ superconductivity and spin orbit coupling in quasi-two dimensional materials carried out here is additionally motivated by the very recent discovery of the supercurrent diode effect\cite{Ando2020}, for which the critical current along opposite directions differ. As it has been argued in recent theoretical papers the observed effect can be directly related to the non-zero momentum pairing which appears in the helical state\cite{Yuan2022,Daido2022,He_2022}. The $extended$-$s$ $wave$ pairing is a good candidate for the description of the paired state in many of the two dimensional materials some of which additionally are characterized by a significant spin-orbit coupling. Therefore, it is very important to theoretically analyze the interplay between this particular pairing symmetry and the SOC in the context of non-zero momentum pairing.

In this work, we apply a single-band lattice model with the nearest-neighbor real space pairing term in order to analyze the general features related with the interplay between $\mathbf{k}$-dependent gap and the spin-orbit coupling in the presence of the in-plane external magnetic field. First, we show that the $extended$ $s$-$wave$ symmetry pairing is the most probable candidate to be realized in the low carrier concentration regime. Moreover, it leads to the dome like behavior of the critical temperature as a function of chemical potential, what is especially interesting in the context of the superconducting LaAlO$_3$/SrTiO$_3$ interfaces. Moreover, as we show the van Hove singularity which resides close to the bottoms of the band and is induced by the spin-orbit coupling enhances the superconducting state leading to a clear $T_c$ peak. Finally, we analyze the non-zero momentum pairing induced by the Fermi wave vector mismatch of the helical bands in the presence of the in-plane magnetic field. As we show in the considered system the helical state is relatively robust and becomes stable as soon as the magnetic field is present, which potentially increases the possibility of its detection. Moreover, the $C_4$ symmetry breaking resulting from the in-plane magnetic field induces a small $d$-$wave$ and $p$-$wave$ contributions leading to a mixed parity of the resulting helical state. 



\section{Model and method}
\begin{figure}[!t]
 \centering
 \includegraphics[width=0.5\textwidth]{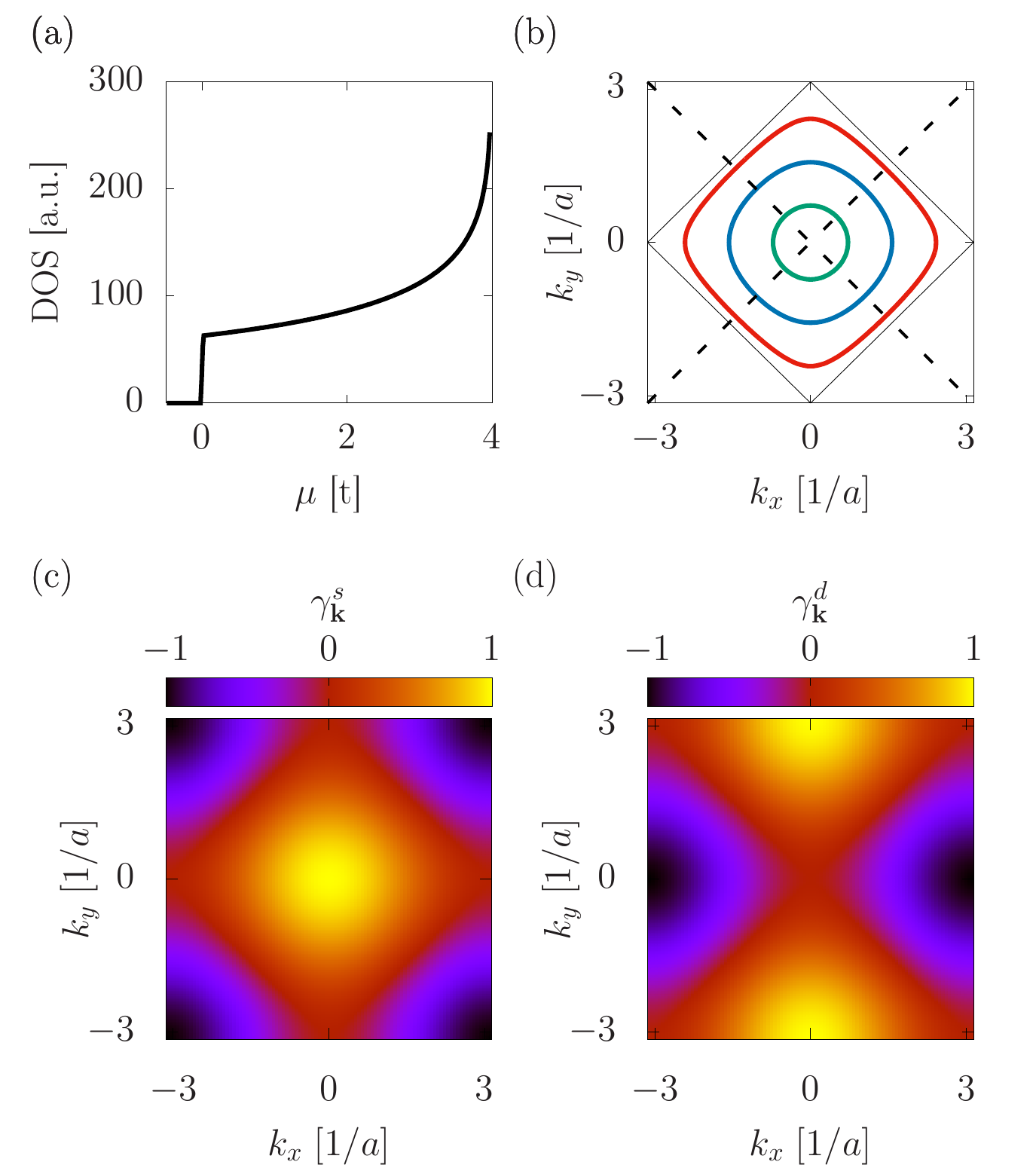}
 \caption{(a) The density of states as a function of chemical potential for the model with $\alpha=0$; (b) The Fermi surfaces for three selected values of the chemical potential: $\mu=0.5\;t$ (green), $\mu=2.0\;t$ (blue), and $\mu=3.5\;t$ (red). The black solid and dashed lines correspond to the nodal lines of the $extended$ $s$-$wave$ and $d$-$wave$ symmetries, respectively. In (c) and (d) we show the momentum dependant symmetry factors corresponding to the $extended$ $s$-$wave$ and $d$-$wave$ pairings, respectively.}
 \label{fig:dos}
\end{figure}

We consider the single-band model with real-space pairing and spin-orbit coupling of the form
\begin{equation}
 \hat{H}=\hat{H}_0+\hat{H}_{SC}, 
 \label{eq:Hamiltonian_start}
\end{equation}
where $\hat{H}_0$ is the single-particle part containing electron hopping, Rashba-type spin-orbit coupling, and interaction between electron spin and the external magnetic field
\begin{equation}
\begin{split}
 \hat{H}_0&=\sum_{\mathbf{k}}(\epsilon_{\mathbf{k}}-\mu)\hat{c}^{\dagger}_{\mathbf{k}\sigma}\hat{c}_{\mathbf{k}\sigma}+\alpha\sum_{\mathbf{k}}\mathbf{g}(\mathbf{k})\cdot \mathbf{S}(\mathbf{k})\\
 &+\mu_{B}g\sum_{\mathbf{k}}\mathbf{B}\cdot\mathbf{S}(\mathbf{k}), 
 \end{split}
 \label{eq:Hamiltonian_0}
\end{equation}
where $\epsilon_{\mathbf{k}}$ is the dispersion relation, $\mu$ is the chemical potential and
\begin{equation}
    \mathbf{S}(\mathbf{k})=\sum_{\mathbf{k}\sigma\sigma'}\pmb{\sigma}_{\sigma\sigma'}\hat{c}^{\dagger}_{\mathbf{k}\sigma}\hat{c}_{\mathbf{k}\sigma'},
    \label{eq:S_vector}
\end{equation}
is the spin operator with $\pmb{\sigma}$ being the Pauli matrices and $\mathbf{g}(\mathbf{k})$ being the vector characterizing the structure of the spin-orbit coupling in the noncentrosymmetric system,
\begin{equation}
    \mathbf{g}(\mathbf{k})=\bigg(-\frac{\partial\epsilon(\mathbf{k})}{\partial k_y}, \frac{\partial\epsilon(\mathbf{k})}{\partial k_x}, 0\bigg).
\end{equation}
The last term in Eq. (\ref{eq:Hamiltonian_0}) introduces the interaction of electron spin with the external magnetic filed $\mathbf{B}=(B_x,B_y,B_z)$ where $\mu_B$ is the Bohr magneton. We focus on the typical situation of square lattice system with non-zero hopping integrals between the nearest neighbors only. The resulting dispersion relation reads
\begin{equation}
    \epsilon_{\mathbf{k}}=4t-2t(\cos k_x + \cos k_y),
\end{equation}
where $t$ is the electron hopping energy taken as $t=0.1\;$eV. In the following section all the results are going to be presented in the units of $t$. For the sake of clarity the quasi momentum is expressed in the units of $1/a$, where $a$ is the lattice constant. In the above equation we additionally include the onsite (atomic) energy term, $4t$, which only shifts the resulting band so that the bottom of the band lies at zero energy. The density of states and Fermi surfaces for three selected values of the chemical potential of such model are provided in Fig. \ref{fig:dos} (a) and (b).

The real-space pairing term responsible for the appearance of the superconducting phase has the following form
\begin{equation}
\begin{split}
 \hat{H}_{SC}&=-J\sum_{i<j,\;\sigma}\hat{c}^{\dagger}_{il\sigma}\hat{c}^{\dagger}_{jl\bar{\sigma}}\hat{c}_{il\bar{\sigma}}\hat{c}_{jl\sigma},
 \end{split}
 \label{eq:Hamiltonian_pairing}
\end{equation}
where we carry out the summation only over the nearest-neighbor bonds ($i<j$). Alternatively, one could consider an onsite real-space pairing scenario which has been discussed in the context of the so-called negative $U$ Hubbard model\cite{Moreo1991} in which isotropic $s$-$wave$ superconducting gap is possible (cf. Appendix A). However, such pairing symmetry is much less likely to appear in the low-carrier concentration regime as shown in the following section.

After carrying out the Hartree-Fock-BCS approximation for the pairing term and transformation to the reciprocal space with the possibility of non-zero momentum pairing one obtains
\begin{equation}
\begin{split}
    \hat{H}_{SC}&=\frac{1}{2}\sum_{k\sigma}\Big(\Delta^{\bar{\sigma}\sigma}_{\mathbf{k}\mathbf{Q}}\;\hat{c}^{\dagger}_{\mathbf{k}\sigma}\hat{c}^{\dagger}_{-\mathbf{k+Q}\bar{\sigma}} + (\Delta^{\bar{\sigma}\sigma}_{\mathbf{k}\mathbf{Q}})^{\star}\;\hat{c}_{-\mathbf{k+Q}\bar{\sigma}}\hat{c}_{\mathbf{k}\sigma}\Big)\\
    &+\frac{1}{J}\sum_{i<j}|\Delta^{\bar{\sigma}\sigma}_{ji}|,
\end{split}
\label{eq:Hamiltonian_pairing_rec_space}
\end{equation}
where
\begin{equation}
    \Delta^{\bar{\sigma}\sigma}_{\mathbf{k}\mathbf{Q}}=\sum_{i(j)}e^{i\mathbf{k}(\mathbf{R}_i-\mathbf{R}_j)}\tilde{\Delta}^{\bar{\sigma}\sigma}_{ji\mathbf{Q}}.
    \label{eq:delta_kQ}
\end{equation}
The summation above runs over the four nearest neighbors of the lattice site $j$ and 
\begin{equation}
\begin{split}
    \tilde{\Delta}^{\bar{\sigma}\sigma}_{ji\mathbf{Q}}&=-J\sum_{\mathbf{k}}e^{-i\mathbf{k}(\mathbf{R}_i-\mathbf{R}_j)}\langle\hat{c}_{-\mathbf{k+Q}\bar{\sigma}}\hat{c}_{\mathbf{k}\sigma}\rangle,
\end{split}
\label{eq:delta_ijQ}
\end{equation}
where $\mathbf{Q}$ is the total momentum of the Cooper pairs. It should be noted that the gap amplitude $\tilde{\Delta}^{\bar{\sigma}\sigma}_{ji\mathbf{Q}}$-parameter depends on the $\mathbf{Q}$ vector in an implicit manner though the expectation value $\langle\hat{c}_{-\mathbf{k+Q}\bar{\sigma}}\hat{c}_{\mathbf{k}\sigma}\rangle$. However, it is not modulated in real space, in contradiction to the true real space pairing amplitude which has the form
\begin{equation}
\begin{split}
    \Delta^{\bar{\sigma}\sigma}_{ji\mathbf{Q}}&=-J\;\langle\hat{c}_{j\bar{\sigma}}\hat{c}_{i\sigma}\rangle\\
    &=-J\;e^{-i\mathbf{R}_j\mathbf{Q}}\sum_{\mathbf{k}}e^{-i\mathbf{k}(\mathbf{R}_i-\mathbf{R}_j)}\langle\hat{c}_{-\mathbf{k+Q}\bar{\sigma}}\hat{c}_{\mathbf{k}\sigma}\rangle\\
    &=e^{-i\mathbf{R}_j\mathbf{Q}}\tilde{\Delta}^{\bar{\sigma}\sigma}_{ji\mathbf{Q}}.
\end{split}
    \label{delta_ij}
\end{equation}
Using Eqs. (\ref{eq:Hamiltonian_0}) and (\ref{eq:Hamiltonian_pairing_rec_space}) we can write down the full Hamiltonian of our system in the matrix form
\begin{equation}
\begin{split}
 \hat{H}&=\sum_{\mathbf{k}}\mathbf{\hat{f}}^{\dagger}
_{\mathbf{k}}\mathbf{\hat{H}}_{\mathbf{k}}\mathbf{\hat{f}}_{\mathbf{k}}+\frac{1}{2}\sum_{\mathbf{k}}\Big(\epsilon_{-\mathbf{k+Q}\uparrow}+\epsilon_{-\mathbf{k+Q}\downarrow}\Big)\\
&+\frac{1}{J}\sum_{i<j}|\Delta^{\bar{\sigma}\sigma}_{ji}|^2
\end{split}
\label{eq:Hamiltonian_final}
\end{equation}
where we have introduced the four-component composite operators
\begin{equation}
 \mathbf{\hat{f}}^{\dagger}_{\mathbf{k}}\equiv(\hat{c}^{\dagger}_{\mathbf
{k}\uparrow},\hat{c}^{\dagger}_{\mathbf{k}\downarrow},\hat{c}_{\mathbf
{-k+Q}\uparrow},\hat{c}_{\mathbf{-k+Q}\downarrow})\;,
\end{equation}
and the Hamiltonian matrix is the following
\begin{equation}
\mathbf{\hat{H}}_{\mathbf{k}}=\frac{1}{2}\left(\begin{array}{cccc}
 \epsilon_{\mathbf{k}\uparrow} & \epsilon_{\mathbf{k}\uparrow\downarrow} & 0 & \Delta^{\downarrow\uparrow}_{\mathbf{k}\mathbf{Q}} \\
\epsilon_{\mathbf{k}\downarrow\uparrow} & \epsilon_{\mathbf{k}\downarrow} & \Delta^{\uparrow\downarrow}_{\mathbf{k}\mathbf{Q}} & 0 \\
0 & \Big(\Delta^{\uparrow\downarrow}_{\mathbf{k}\mathbf{Q}}\Big)^{\star} & -\epsilon_{\mathbf{-k+Q}\uparrow} & -\epsilon_{\mathbf{-k+Q}\uparrow\downarrow} \\
\Big(\Delta^{\downarrow\uparrow}_{\mathbf{k}\mathbf{Q}}\Big)^{\star} & 0 & -\epsilon_{\mathbf{-k+Q}\downarrow\uparrow} & -\epsilon_{\mathbf{-k+Q}\downarrow} \\
\end{array} \right)\;.
\label{eq:Hamiltonian_matrix}
\end{equation}
In the above, we have introduced the following notation
\begin{equation}
    \epsilon_{\mathbf{k}\sigma}=\epsilon_{\mathbf{k}}+\sigma\frac{\mu_B g}{2}B_z-\mu,
\end{equation}
\begin{equation}
    \epsilon_{\mathbf{k}\uparrow\downarrow}=\alpha\Big(g_x(\mathbf{k})-ig_y(\mathbf{k})\Big)+\frac{\mu_B g}{2}(B_x-iB_y),
\end{equation}
\begin{equation}
    \epsilon_{\mathbf{k}\downarrow\uparrow}=\alpha\Big(g_x(\mathbf{k})+ig_y(\mathbf{k})\Big)+\frac{\mu_B g}{2}(B_x+iB_y).
\end{equation}
One can derive the self-consistent equations for the superconducting gap in a standard manner by writing down the gap equation (\ref{eq:delta_ijQ}) in the new basis which results from the diagonalization of the matrix Hamiltonian given by Eq. (\ref{eq:Hamiltonian_matrix}). 

It is instructive to look at the pairing term $\hat{H}_{SC}$ represented in terms of the creation and annihilation operators corresponding to states from the two helical bands created due to the spin-orbit coupling. In such representation $\hat{H}_{SC}$ reads
\begin{equation}
\begin{split}
    \hat{H}_{SC}&=\frac{1}{2}\sum_{kl}\Big(\tilde{\Delta}^{(l)}_{\mathbf{k}\mathbf{Q}}\;\hat{\alpha}^{\dagger}_{\mathbf{k}l}\hat{\alpha}^{\dagger}_{-\mathbf{k+Q}l} + (\tilde{\Delta}^{(l)}_{\mathbf{k}\mathbf{Q}})^{\star}\;\hat{\alpha}_{-\mathbf{k+Q}l}\hat{\alpha}_{\mathbf{k}l}\Big)\\
    &+\frac{1}{J}\sum_{i<j}|\Delta^{\bar{\sigma}\sigma}_{ji}|,
    \label{eq:Hamiltonian_pairing_rec_helical}
\end{split}
\end{equation}
where $l=\pm$ correspond to the two helical bands with dispersion relations
\begin{equation}
    \begin{split}
        \tilde{\epsilon}_{\mathbf{k}\pm}&=\epsilon_{\mathbf{k}}\pm\alpha|\mathbf{\tilde{g}}(k)|,\\
    \end{split}
    \label{eq:helical_bands}
\end{equation}
where the $\mathbf{\tilde{g}}(\mathbf{k})$ vector contains both the SOC and the external in-plane magnetic field contributions and has the form
\begin{equation}
\begin{split}
    \mathbf{\tilde{g}}(\mathbf{k})&=\big(\tilde{g}_x(\mathbf{k}),\quad\tilde{g}_y(\mathbf{k})\big)\\
    &=\Big(\alpha\;g_x(\mathbf{k})+\frac{\mu_B\;g}{2} B_x,\quad\alpha\;g_y(\mathbf{k})+\frac{\mu_B\;g}{2} B_y \Big).
\end{split}
\label{eq:g_tilda}
\end{equation}
The SC gap created in the helical bands can now be expressed in terms of the original spin-resolved gap amplitudes in the following manner 
\begin{equation}
    \tilde{\Delta}^{(\pm)}_{\mathbf{k}\mathbf{Q}}=\pm\Big(G_{\mathbf{k}}^{\star}\Delta^{\uparrow\downarrow}_{\mathbf{k}\mathbf{Q}}+G_{\mathbf{-k+Q}}^{\star}\Delta^{\downarrow\uparrow}_{\mathbf{k}\mathbf{Q}}\Big),
    \label{eq:delta_helical}
\end{equation}
where
\begin{equation}
    G_{\mathbf{k}}^{\star}=\frac{1}{|\mathbf{\tilde{g}(k)}|}\Big(\tilde{g}_x(\mathbf{k})+i \tilde{g}_y(\mathbf{k})\Big).
    \label{eq:G_factor}
\end{equation}
In Appendix B we describe in more detail how to derive eqs. \ref{eq:Hamiltonian_pairing_rec_helical} and \ref{eq:delta_helical}.
From Eq. (\ref{eq:Hamiltonian_pairing_rec_helical}) one can see that the SC gaps in the two helical bands are equal with respect to the absolute values but have opposite signs which means that we are dealing with a $(\pm)$-type of pairing. Also, it should be noted that in the absence of the external magnetic filed ($\mathbf{B}=0$ and $\mathbf{Q}=0$), we have $G_{-\mathbf{k}}^{\star}=-G_{\mathbf{k}}^{\star}$ and the SC gaps in the helical bands can be rewritten in the following form
\begin{equation}
    \tilde{\Delta}^{(\pm)}_{\mathbf{k}}=\pm G_{\mathbf{k}}^{\star}\Big(\Delta^{\uparrow\downarrow}_{\mathbf{k}}-\Delta^{\downarrow\uparrow}_{\mathbf{k}}\Big),
    \label{eq:delta_helical_zero_mementum}
\end{equation}
from which one can see, that the superconducting pairing in the two helical bands results in a straightforward manner from the spin singlet pairing in the original spin-resolved basis. However, the $\mathbf{k}$-dependence of the gap is modified by the $G^{\star}_{\mathbf{k}}$ function which contains the contribution form the spin-orbit coupling. Since $\Delta^{\sigma\sigma'}_{\mathbf{k}}$ is even, the overall $\mathbf{k}$-dependence of $\tilde{\Delta}^{(\pm)}_{\mathbf{k}}$ is odd, as it should be since now we have intra-helical band pairing. After the application of the external magnetic field one obtains, $G^{\star}_{\mathbf{k}}\neq G^{\star}_{-\mathbf{k}}$ and the weights with which $\Delta^{\uparrow\downarrow}$ and $\Delta^{\downarrow\uparrow}$ contribute to the resulting $\tilde{\Delta}^{(\pm)}$ are different with respect to their absolute value. 

Due to the presence of both spin-orbit coupling and the external magnetic field apart from the spin-singlet component one also obtains a spin-triplet contribution to the pairing in spite of the fact that the original pairing mechanism has a pure spin-singlet character [cf. Eq. (\ref{eq:Hamiltonian_pairing})]. In such circumstances a mixture of different pairing symmetries may appear in the superconducting state. It is convenient to introduce the following symmetry resolved gap amplitudes
\begin{equation}
\begin{split}
    \Delta^s_s&=\big(\tilde{\Delta}^{\uparrow\downarrow}_{10}+\tilde{\Delta}^{\uparrow\downarrow}_{10}+\tilde{\Delta}^{\uparrow\downarrow}_{01}+\tilde{\Delta}^{\uparrow\downarrow}_{01}\\
    &-\Delta^{\downarrow\uparrow}_{10}-\Delta^{\downarrow\uparrow}_{10}-\Delta^{\downarrow\uparrow}_{01}-\Delta^{\downarrow\uparrow}_{01}\big)/8,\\
    \Delta^s_d&=\big(\tilde{\Delta}^{\uparrow\downarrow}_{10}+\tilde{\Delta}^{\uparrow\downarrow}_{10}-\tilde{\Delta}^{\uparrow\downarrow}_{01}-\tilde{\Delta}^{\uparrow\downarrow}_{01}\\
    &-\tilde{\Delta}^{\downarrow\uparrow}_{10}-\tilde{\Delta}^{\downarrow\uparrow}_{10}+\tilde{\Delta}^{\downarrow\uparrow}_{01}+\tilde{\Delta}^{\downarrow\uparrow}_{01}\big)/8,\\
    \Delta^t_{p_x}&=\big(\tilde{\Delta}^{\uparrow\downarrow}_{1,0}-\tilde{\Delta}^{\downarrow\uparrow}_{-1,0}+\tilde{\Delta}^{\downarrow\uparrow}_{1,0}-\tilde{\Delta}^{\uparrow\downarrow}_{-10}\big)/4,\\
    \Delta^t_{p_y}&=\big(\tilde{\Delta}^{\uparrow\downarrow}_{0,1}-\tilde{\Delta}^{\downarrow\uparrow}_{0,-1}+\tilde{\Delta}^{\downarrow\uparrow}_{0,1}-\tilde{\Delta}^{\uparrow\downarrow}_{0,-1}\big)/4,
\end{split}
\label{eq:gap_symmetries}
\end{equation}
where $\tilde{\Delta}^{\tilde{\sigma}\sigma}_{\mathbf{R}_{ji}}=\tilde{\Delta}^{\tilde{\sigma}\sigma}_{ji\mathbf{Q}}$, and $\mathbf{R}_{ji}=\mathbf{R}_i-\mathbf{R}_j$ is the vector adjoining the two nearest neighbor lattice sites between which the pairing appears (expressed in the units of the lattice constants, $a$). In the notation for $\tilde{\Delta}^{\tilde{\sigma}\sigma}_{\mathbf{R}_{ji}}$ we have dropped the $\mathbf{Q}$ index for the sake of clarity. The upper index on the left hand side of the equations corresponds to the spin-symmetry ($s$ for singlet, $t$ for triplet), while the lower index corresponds to the real space (or equivalently reciprocal space) symmetry of the superconducting gap ($s$ for $extended$ $s$-$wave$, $d$ for $d$-$wave$, $p_x/p_y$ for the two possibilities of the $p$-$wave$ symmetries). The relation with the $\mathbf{k}$-depended gap amplitudes $\Delta^{\bar{\sigma}\sigma}_{\mathbf{k}\mathbf{Q}}$ appearing in Eqs. (\ref{eq:Hamiltonian_pairing_rec_space}), (\ref{eq:delta_helical}), (\ref{eq:delta_helical_zero_mementum}) is provided in the Appendix C. In the presence of both external magnetic field and spin-orbit coupling all the above pairing amplitudes may become nonzero. However, for $\mathbf{B}=0$ and $\alpha=0$ only $\Delta^s_s$ or $\Delta^s_d$ may have nonzero values which corresponds to pure $extended$ $s$-$wave$ or $d$-$wave$ pairing symmetries for which the $\mathbf{k}$-dependence of the superconducting gap has the form $\Delta_{s}(\mathbf{k})=4\Delta_s^s\gamma^s_{\mathbf{k}}$ or $\Delta_{d}(\mathbf{k})=4\Delta_s^s\gamma^d_{\mathbf{k}}$, respectively. The gamma symmetry factors are the following
\begin{equation}
\begin{split}
    \gamma^s_{\mathbf{k}}&=(\cos k_x + \cos k_y)/2,\\
    \gamma^d_{\mathbf{k}}&=(\cos k_x - \cos k_y)/2. 
\end{split}
\label{eq:symmetry_factors}
\end{equation}
For the sake of completeness those factors are presented in Fig. \ref{fig:dos} (c) and (d) as functions of momentum. As shown in Fig. \ref{fig:dos} (b) for both factors one can define the so-called nodal lines in the reciprocal space for which the superconducting gap closes down. Therefore, if the Fermi surface of our system is at its whole extent in close proximity of the nodal lines of a given pairing symmetry one should expect a suppression of the superconducting phase.

\section{Results}
\begin{figure*}[!t]
 \centering
 \includegraphics[width=1.0\textwidth]{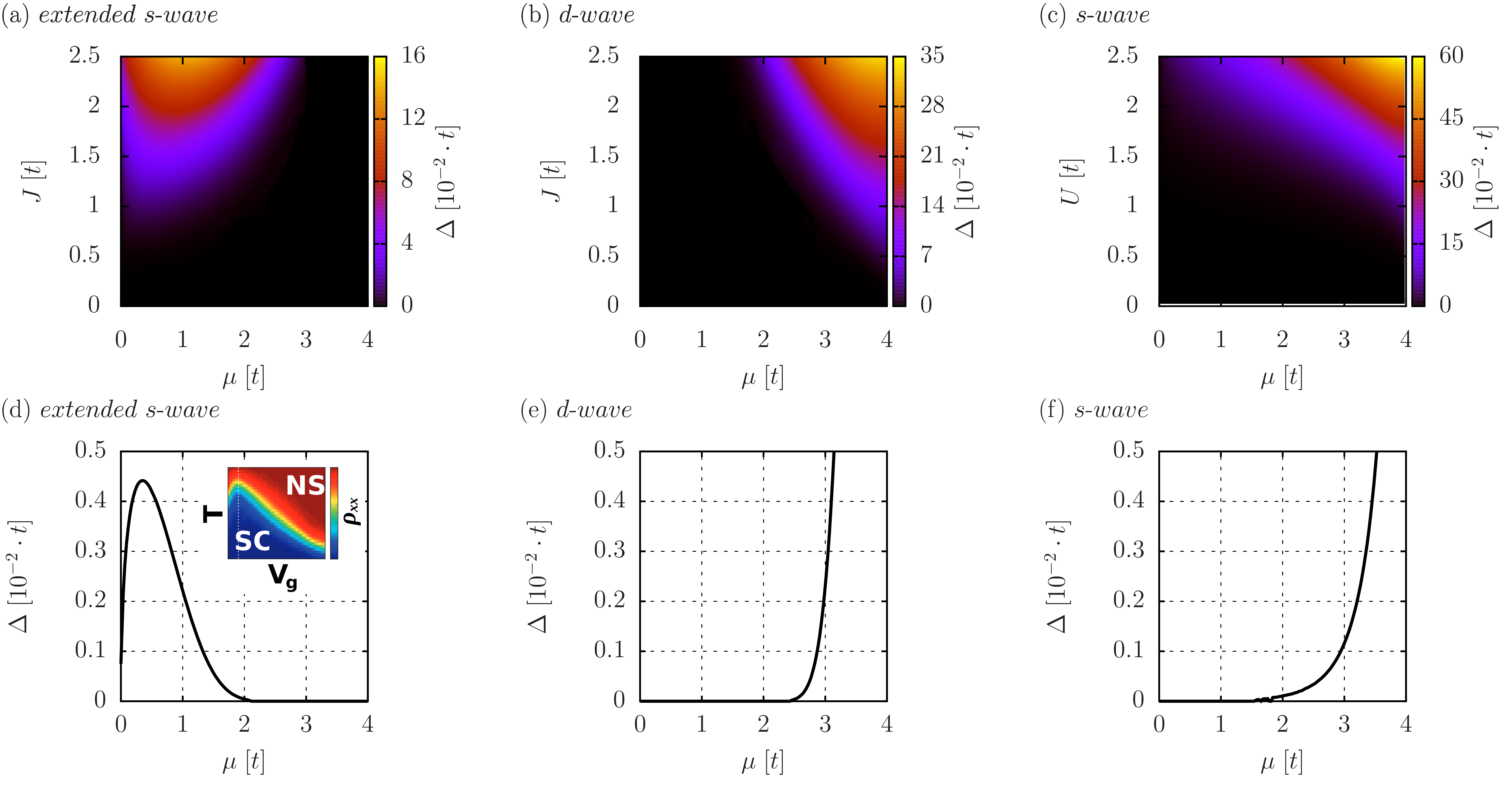}  
 \caption{(a), (b), (c) - The superconducting gap amplitude as a function of chemical potential and superconducting coupling constant $J$ for $\alpha=0$ and $\mathbf{B}=0$ for three different possible symmetries of the real space paired state. The two limiting values $\mu=0$ and $\mu=4\;t$ correspond to the bottom of the band and to half-filled situation, respectively. The two cases of $extended$ $s$-$wave$ and $d$-$wave$ symmetries correspond to the model with nearest neighbor pairing while the $s$-$wave$ symmetry is induced by the onsite pairing term (cf. Appendix A). In (d), (e), and (f) we show the gap amplitudes as a function of chemical potential for selected value of $J=0.8\;t$. Note the appearance of the SC-dome as a function of $\mu$ in the low carrier concentration regime only for the case of $extended$ $s$-$wave$ symmetry of the superconducting gap. In the inset of (d) we provide the experimental result showing the dome like behavior of $T_C$ as a function of gate voltage for the LaAlO$_3$/SrTiO$_3$ interface taken from Ref. \onlinecite{joshua2012universal}.}
 \label{fig:mu_J_phase_diags}
\end{figure*}

Before we analyze the influence of the spin-orbit coupling on the superconducting state and the possibility of the non-zero momentum pairing, we calculate the phase diagram in the ($\mu$, $J$)-plane in the absence of the external magnetic field and for $\alpha=0$. In Fig. \ref{fig:mu_J_phase_diags} we provide the SC gap amplitude for the pairing scenarios corresponding to pure $extended$ $s$-$wave$ and pure $d$-$wave$ symmetries, which $\mathbf{k}$-dependence is determined by the corresponding symmetry factors provided in Eqs. (\ref{eq:symmetry_factors}). For the sake of completeness we also provide analogical figures for the case of the on-site pairing scenario which leads to isotropic $s$-$wave$ symmetry (cf. Appendix A for the explicit form of the paring term in such case). As one could expect the stronger the coupling constants $J$ and $U$ ($U$ defines the onsite pairing strength) the larger the gap in all three cases. However, the region of stability of the SC phase strongly depends on the pairing symmetry. Namely, in the low electron concentration regime (close to the bottom of the band) the $extended$ $s$-$wave$ symmetry dominates while for larger values of $\mu$ (close to the half-filled situation) the $d$-$wave$ or $s$-$wave$ pairing becomes stable depending on the choice of the particular pairing term (inter-site or on-site pairing). Interestingly, in the low electron concentration regime, a dome-like behaviour of the $extended$ $s$-$wave$ pairing amplitude appears as a function of chemical potential for a given value of $J$ [cf. Fig. \ref{fig:mu_J_phase_diags} (d)].

It should be noted that the two-dimensional superconducting system with very low electronic concentration and Fermi energy close to the bottom of the bands is realized in the LaAlO$_3$/SrTiO$_3$ interfaces, where additionally a dome-like behavior of the critical temperature as a function of chemical potential has been reported experimentally\cite{Reyren2007,Rout2017, joshua2012universal,maniv2015strong}. With respect to these experimental findings different gap symmetries have been discussed as well as various physical mechanisms leading to the dome-like superconductivity have been proposed\cite{maniv2015strong,Smink,Trevisan2018,Singh2018gap,Lepori2021,Perroni2021,Citro_2022}. The study presented here, based on a relatively simple single band model, shows that the $extended$ $s$-$wave$ pairing symmetry is the most probable one for the low-electron concentration regime, at least for the case of real-space spin-singlet pairing. Additionally, such choice leads to the appearance of the dome-like behavior of $T_c$ in a natural manner. This may point to the scenario in which the dome-like shape of $T_C$ in LaAlO$_3$/SrTiO$_3$ interfaces as a function of carrier concentration comes not as a result of electron-electron correlation or complex multiband effects but is caused simply by the symmetry of the gap. This would be consistent with the experimental results shown in Refs. \onlinecite{Singh2018,Biscaras} according to which the Lifshitz transition appears at lower critical carrier concentration for the appearance of the superconducting state. In such circumstance, it is tempting to compare the dome-like shape of the gap amplitude obtained here with the experimental result corresponding to the LaAlO$_3$/SrTiO$_3$ interface. Such comparison is shown in the inset of Fig. \ref{fig:mu_J_phase_diags} (d).

This last effect (dome-like behaviour) can be understood as an result of relative distance between the Fermi surfaces (FS) and the nodal lines of the $extended$ $s$-$wave$ symmetry. With increasing $\mu$ the SC amplitude rises rapidly at first due to the jump of the density of states which appears at the bottom of the band (cf. Fig. \ref{fig:dos}a). Than, with further increase of the chemical potential while the Fermi surface expands, it is getting closer and closer to the nodal lines (cf. Fig. \ref{fig:dos}), where the gap closes strictly due to the symmetry of the $extended$ $s$-$wave$ pairing. The close proximity of the Fermi surface to the nodal lines suppresses the gap at the FS and leads to the decrease of the SC pairing amplitude with further increase of $\mu$. In between those two regimes the optimal carrier concentration is located for which the maximal SC pairing amplitude appears.



\begin{figure}[!t]
 \centering
 \includegraphics[width=0.5\textwidth]{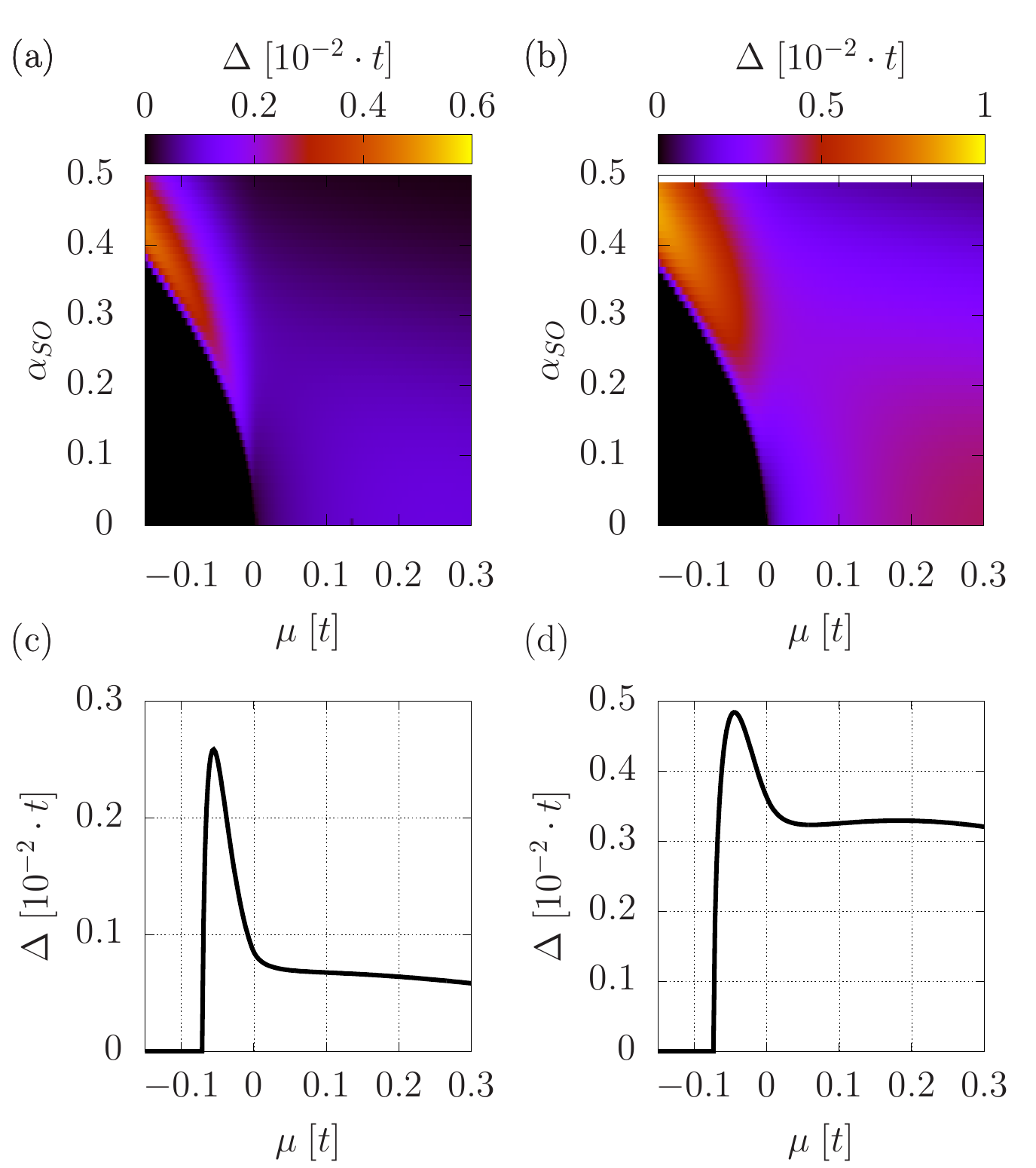}
 \caption{(a) and (b) - The superconducting gap as a function of both Fermi energy and the SO coupling energy for two different values of the real space Cooper pairing constant $J=0.6\;t$ and $J=0.8\;t$, respectively; (c) and (d) - The superconducting gap as a function of the Fermi energy for $\alpha_{SO}=0.25$ as well as for $J=0.6\;t$ and $J=0.8\;t$, respectively.}
 \label{fig:alp_mu_phase_diags}
\end{figure}

\begin{figure}
 \centering
 \includegraphics[width=0.5\textwidth]{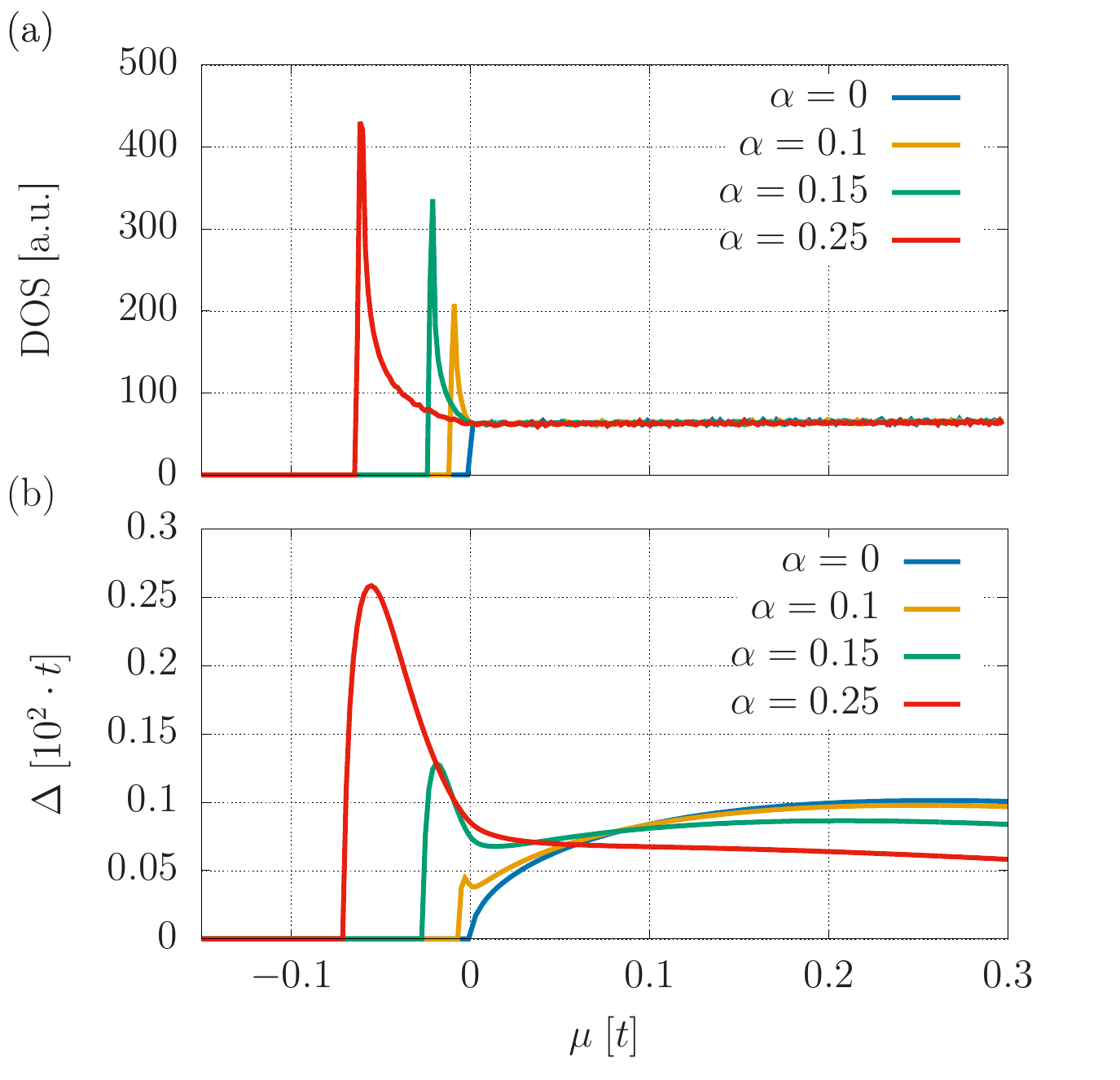}
 \caption{(a) The density of states as a function of the Fermi energy for the four selected values of the $\alpha$ parameter. Note the creation of a van Hove singularity at the bottom of the band after the inclusion of the SOC.; (b) Superconducting gap as a function of Fermi energy for the same four values of the $\alpha$ parameter as in (a) and for a selected value of $J$, $J=0.6\;t$. Note that the van Hove singularity seen in (a) leads to a peak in the pairing gap amplitude in (b).}
 \label{fig:SC_peak_dos}
\end{figure}

Next, we analyze the influence of the spin-orbit coupling on the SC state. In Fig. \ref{fig:alp_mu_phase_diags} we show how the $extended$ $s$-$wave$ SC gap amplitude evolves as we increase the $\alpha$ parameter which determines the strength of the SOC. As one can see for $\alpha>0$ the minimal critical chemical potential for the appearance of the SC state is decreased and a significant enhancement of the SC gap appears in the regime $\mu<0\;$. Those two effects are the consequence of the modifications of the electronic structure introduced by the SOC. Namely, as one increases $\alpha$, the bottom of the created helical bands are decreased below $E=0\;$. This causes the decrease of the lower critical $\mu$ for the appearance of SC. Moreover, in such case the bottoms of the subbands are shifted towards $\mathbf{k}\neq 0$ and a van Hove singularity is created which enhances the SC state. Nevertheless, for a given value of $\alpha$, the effect of the SC amplitude enhancement is more visible for lower values of $J$ [cf. Fig. \ref{fig:alp_mu_phase_diags} (c) and (d)]. In Fig. \ref{fig:SC_peak_dos} we show explicitly that the peak of the SC gap amplitude corresponds to the van Hove singularity which appears in the density of states as a function of $\mu$. As one can see, for $\alpha=0$ when the van Hove singularity is absent at the bottom of the band also the mentioned peak of the gap disappears. It should be noted that the enhancement of superconductivity induced by the van Hove singularity has been discussed in Refs. \onlinecite{Bok2012, Cappelluti2007} in the context of different systems and/or other types of pairing symmetries than the one analyzed here.

\begin{figure}
 \centering
 \includegraphics[width=0.5\textwidth]{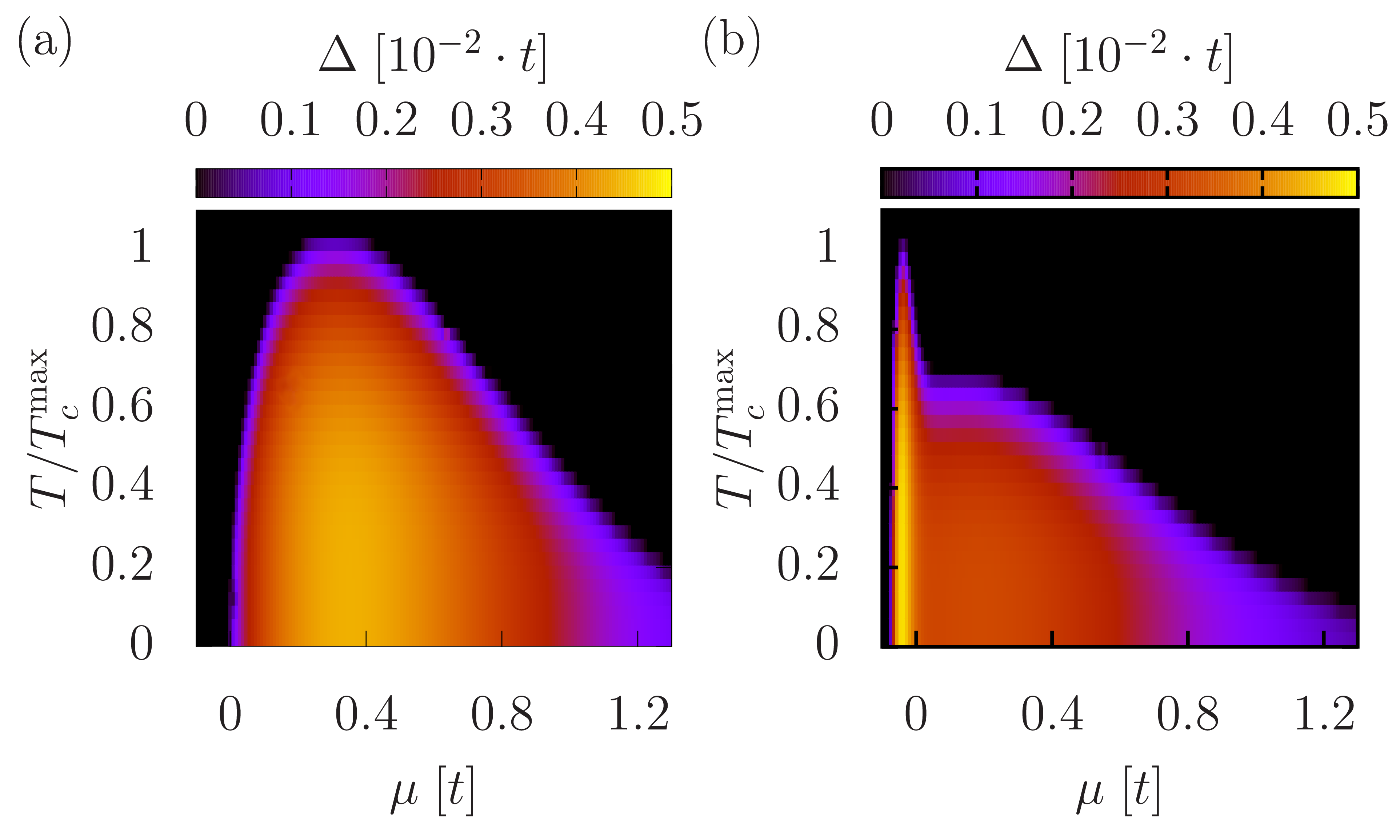}
 \caption{The superconducting gap as a function of Fermi energy and temperature for two selected values of spin orbit coupling $\alpha=0$ and $\alpha=0.25$, which correspond to (a) and (b), respectively. Both diagrams have been obtained for $J=0.8\;t$.}
 \label{fig:temp_mu_phase_diags}
\end{figure}

\begin{figure}
 \centering
 \includegraphics[width=0.5\textwidth]{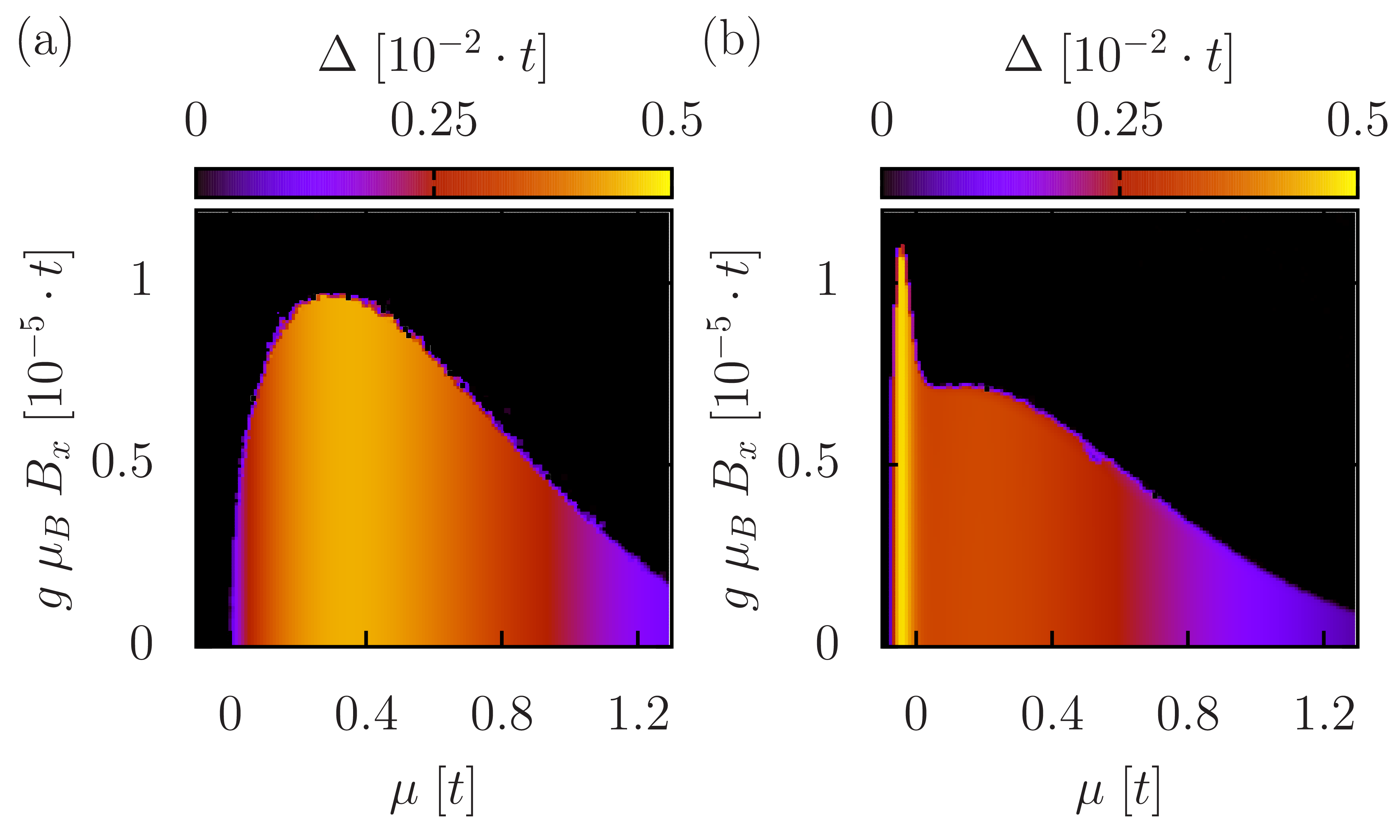}
 \caption{The superconducting gap as a function of Fermi energy and in-plane external magnetic field for two selected values of spin orbit coupling $\alpha=0$ and $\alpha=0.25$, which correspond to (a) and (b), respectively. Both diagrams have been obtained for $J=0.8\;t$.}
 \label{fig:Bx_mu_phase_diags}
\end{figure}

For the sake of completeness, in Fig. \ref{fig:temp_mu_phase_diags} we show the results for non-zero temperatures both for $\alpha=0$ and $\alpha\neq 0$. As one can see the dome like behaviour seen in Fig. \ref{fig:mu_J_phase_diags} (d) is reflected by the critical temperature for $\alpha=0$, while for $\alpha\neq 0$ a very narrow critical temperature peak appears induced by the van Hove singularity. Furthermore, in Fig. \ref{fig:Bx_mu_phase_diags} we show similar results but for the case of non-zero external magnetic field in the $(\mu,B_x)$ plane. Again, both effects seen earlier (dome-like behavior and SC peak) are also reconstructed in the chemical potential dependencies of the critical magnetic field. However, in this particular result we have not yet taken into account the possibility of non-zero momentum pairing, what is done in the following part of this Section.

\begin{figure}
 \centering
 \includegraphics[width=0.5\textwidth]{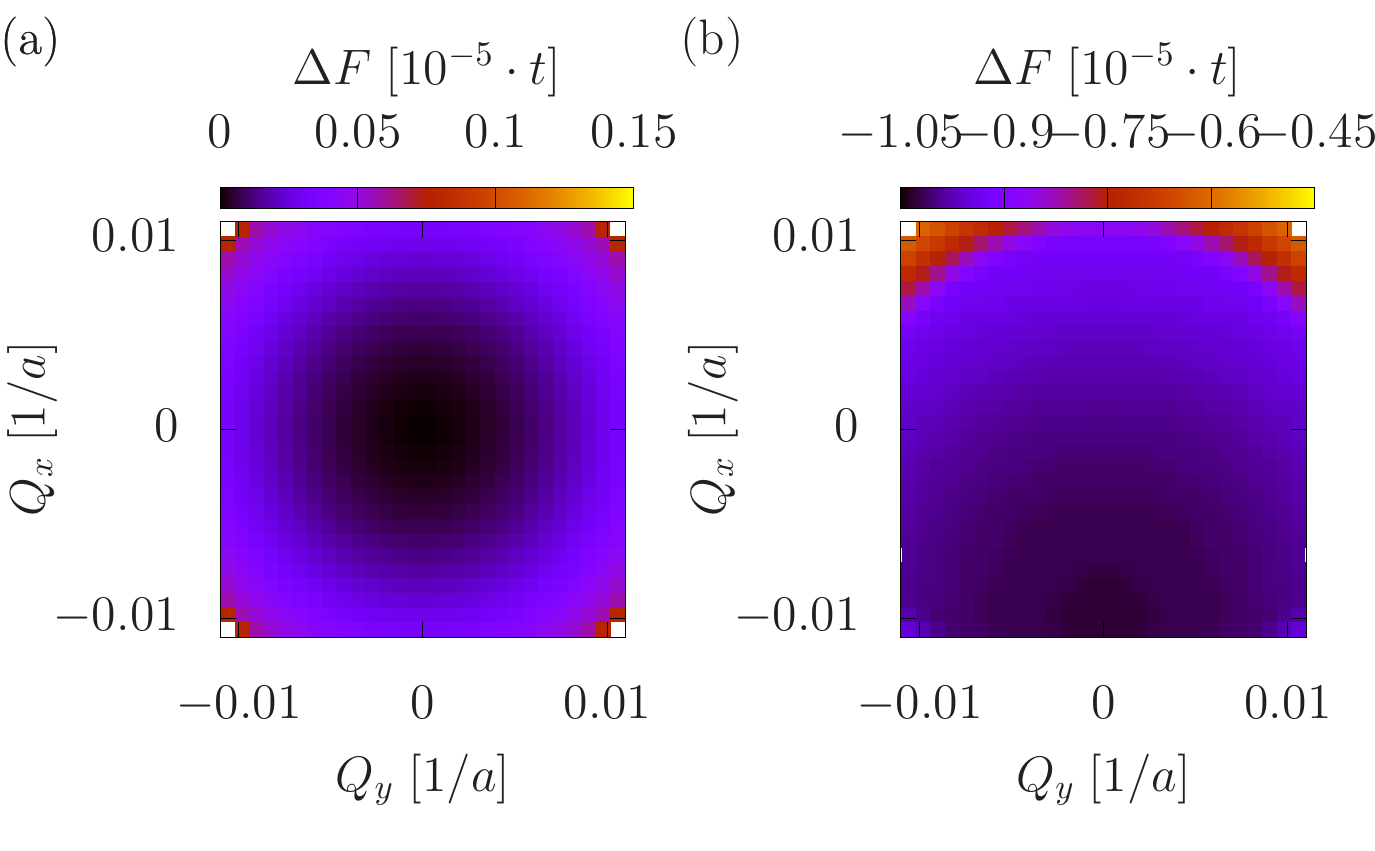}
 \caption{The Cooper pair momentum dependence of the free energy in the superconducting state for $\mu=0.4\;t$, $\alpha=0.25$, $J=0.8\;t$, and two selected values of the in-plane magnetic field $g\mu_B B_x=0.0$ (a) and $g\mu_B B_x=0.17\cdot 10^{-5}\;t$ (b). Note the appearance of the free energy minimum for $\mathbf{Q}\neq 0$ in (b) which means that a non-zero Cooper pair momentum takes place.}
 \label{fig:QxQy}
\end{figure}

In Fig. \ref{fig:QxQy} we show the calculated free energy in the superconducting state as a function of the Cooper pair momentum for selected value of chemical potential $\mu=0.4\;t$ and in the presence of external magnetic field directed towards the $x$-axis. As one can see for $B_x=0$ a free energy minimum appears for $Q_x=Q_y=0$ which means that standard paired state is stable with zero Cooper pair momentum. As one increases the magnetic field to $g\mu_B B_x=0.173\cdot 10^{-5}\;t$, the free energy minimum moves along the $Q_y$-axis and is located at $Q_x=0$ and $Q_y<0$, meaning that the stable paired state is created with the non-zero Cooper pair momentum perpendicular to the magnetic field, which is a characteristic feature of the helical state. The non-zero $\mathbf{Q}$-vector compensates the Fermi-wave-vector mismatch which is created as a consequence of both spin-orbit coupling and the presence of the external magnetic field.  

\begin{figure}
 \centering
 \includegraphics[width=0.5\textwidth]{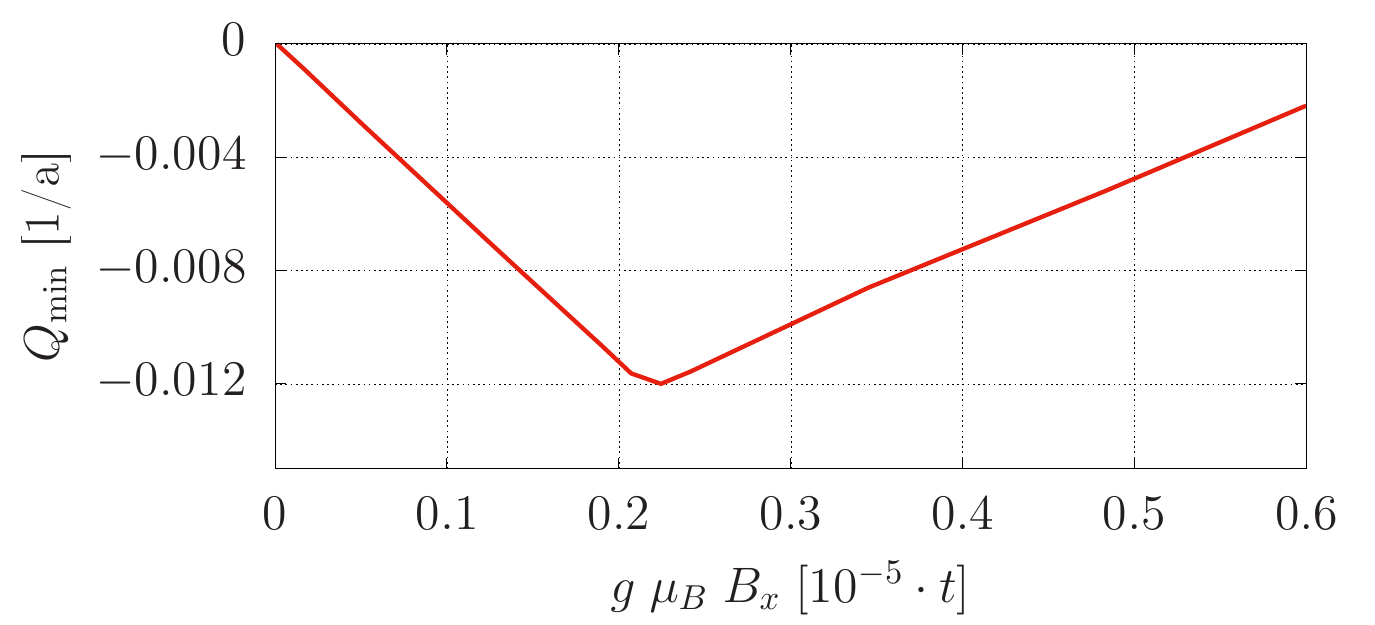}
 \caption{Cooper pair momentum which minimizes the free energy, as a function of the external magnetic field for $J=0.8\;t$, $\mu=0.4\;t$, and $\alpha=0.25$.}
 \label{fig:Qopt_Bx}
\end{figure}

In Fig. \ref{fig:Qopt_Bx} we show how the $y$-axis component of the Cooper pair momentum, corresponding to the free energy minimum, changes as one increases the in-plane magnetic field. It should be noted that in the considered case the helical state is very robust and the paired state can survive up to much larger magnetic fields with respect to the corresponding zero momentum paired state. Such result is consistent with the theoretical analysis of the helical state stability in the spin-orbit coupled two-dimensional ultracold Fermi gases\cite{Zheng2014} where, however, an onsite pairing mechanism has been considered. This is in contradiction to the standard Zeeman splitting induced FFLO state which stability regime is relatively narrow\cite{Matsuda2007}. Interestingly, for the model of inter-site paired state considered here the helical phase becomes stable as soon as the in-plane magnetic field is present (for low $B$).

\begin{figure*}
 \centering
 \includegraphics[width=1.0\textwidth]{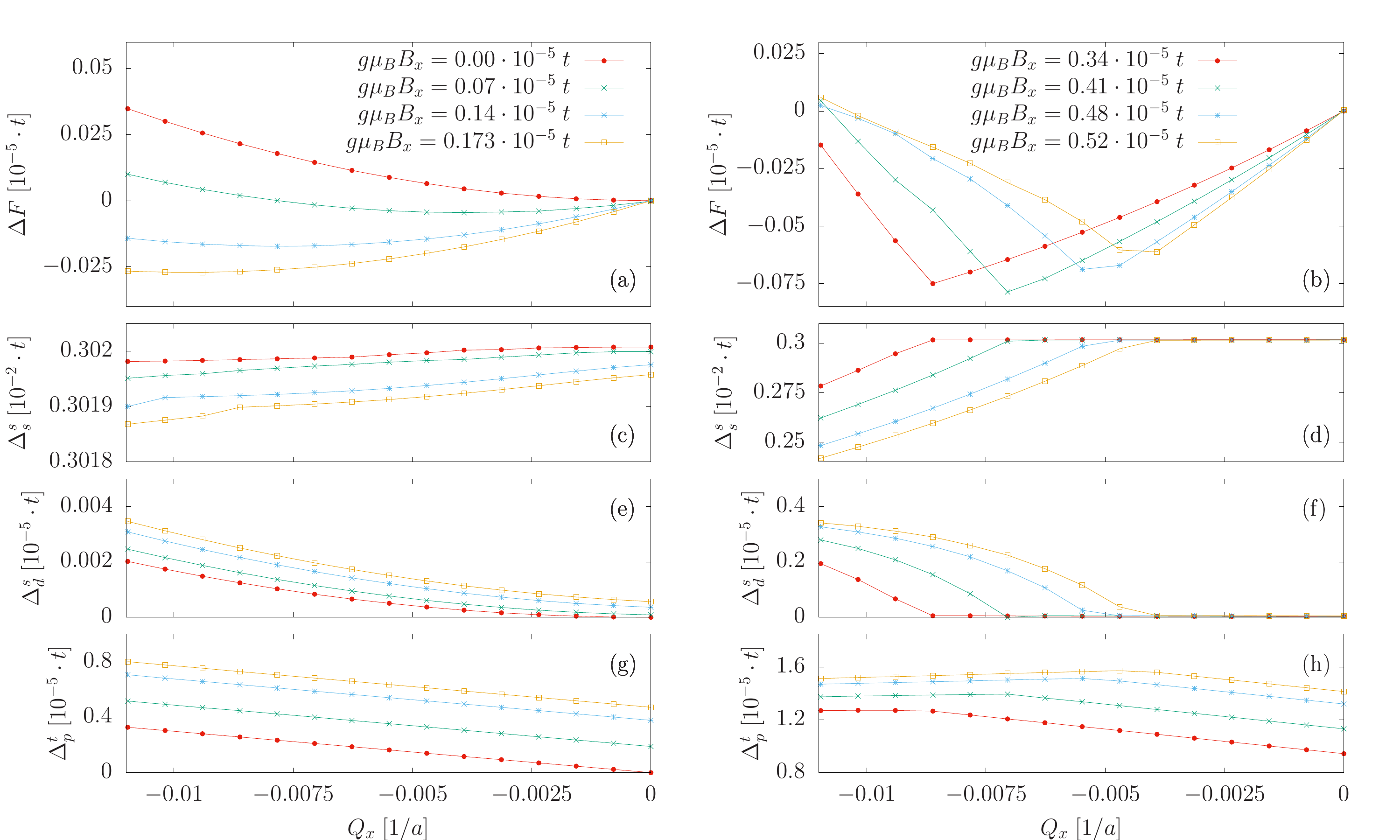}
 \caption{(a,b) The free energy of the system as a function of Cooper pair momentum for selected values of external magnetic field directed along the $x$-axis; (c,d) The $extended$ $s$-$wave$ spin-singlet amplitude of the pairing; (e,f) The $d$-$wave$ spin-singlet amplitude of the pairing; (g,h) The $p_x$-$wave$ spin-triplet pairing amplitude.}
 \label{fig:big}
\end{figure*}

It should be noted that two regions of the helical phase appearance can be distinguished while increasing the in-plane magnetic field (cf. Fig. \ref{fig:Qopt_Bx}). In the first region the absolute value of the Cooper pair momentum increases linearly as a function of the magnetic field. In this region the $Q_y$-dependence of the free energy is approximately parabolic for given $B_x$ and the $extended$ $s$-$wave$ gap amplitude very weakly depends on $\mathbf{Q}$ as shown in Figs. \ref{fig:big} (a) and (c). Also, very small contributions of the $d$-$wave$ and $p_x$-$wave$ pairing appear, which are few orders of magnitude smaller than the corresponding dominant $extended$ $s$-$wave$ amplitude (cf. Figs. \ref{fig:big} e, g).

Beginning from the value of $g\mu_B B_x\approx 0.24\cdot 10^{-5}\;t$, the Cooper pair momentum starts to decrease as one increases the magnetic field. In this region, for a given $B_x$, a sharp minimum appears in the $Q_y$-dependence of the free energy which additionally corresponds to the sudden drop of the $extended$ $s$-$wave$ gap amplitude as shown in Figs. \ref{fig:big} (b) and (d)]. This drop is related with a significant increase of the $d$-$wave$ contribution to the superconducting state as shown in Fig. \ref{fig:big} (f). It should be noted that in spite of the fact that the Cooper pair momentum as a function of external magnetic field has different character in the first two described regions, the value of the dominant gap amplitude $\Delta_s^s$ which corresponds to the free energy minimum is almost constant.

According to our calculations the physical situation does not change when the direction of the in-plane magnetic field is varied. This is because of the fact that we are focused here at the low-electron concentration regime for which the corresponding Fermi surface is approximately circular and, therefore, the analyzed effects have an isotropic character with respect to the in-plane magnetic field.

One should also note that the non-zero Cooper pair momentum, $\mathbf{Q}$, which is generated in the direction perpendicular to the magnetic field breaks time-reversal and inversion symmetry in the equilibrium state. This in turn can lead to a situation in which the critical currents along and against the $\mathbf{Q}$ vector are different. Such non-reciprocal superconducting features may allow for the realization of an intrinsic supercurrent diode effect (SDE) controlled by the magnetic field with the use of the mentioned transition metal oxide interfaces (e.g. LaAlO$_3$/SrTiO$_3$).

\section{Conclusions}
We have analyzed the general features of the real-space intersite paired state in the presence of the Rashba type spin-orbit coupling and external in-plane magnetic field for a two-dimensional superconductor. According to our study, in the low electron concentration regime the $extended$ $s$-$wave$ gap amplitude is dominant and shows a dome-like behavior as a function of chemical potential, which is a characteristic feature of this particular gap symmetry. After the introduction of the spin-orbit coupling a narrow peak of the superconducting gap amplitude appears around the chemical potential value which corresponds to the van Hove singularity created due to the appearance of the SOC. This effect may be potentially useful in the search of quasi-two dimensional superconductors with higher $T_c$'s.

As expected due to the combined result of the external in-plane magnetic field and the spin-orbit coupling the so-called helical paired state emerges. In the analyzed situation, the non-zero momentum Cooper pairing is much more robust as a function of magnetic field than in the case of a standard Zeeman splitting-induced FFLO state\cite{Matsuda2007}. Also, an interesting feature of the reported state is that the non-zero momentum pairing appears as soon as the external magnetic field is applied. Therefore, there is no lower critical magnetic field for the appearance of the helical state as it was reported in previous studies for different models\cite{Loder2013, Matsuda2007, Kok2019, Zheng2014}. This feature makes the quasi-two dimensional superconducting systems with SOC and $extended$ $s$-$wave$ pairing promising candidates for the experimental observation of the non-zero momentum pairing.

According to our analysis, an unconventional in-plane magnetic field dependence of the Cooper pair momentum appears in the studied systems. With this respect, two regions can be distinguished with different behavior of the $\mathbf{Q}$ vector as a function of $\mathbf{B}$. In both regions the dominant $extended$ $s$-$wave$ gap amplitude remains almost constant. However, the Cooper pair momentum linearly increases with magnetic field in the first region while it decreases in the second (cf. Fig. \ref{fig:Qopt_Bx}). The latter effect comes as a result of the $d$-$wave$ component of the gap amplitude which becomes significant above some particular value of $B_x$.

Even though the model presented here is not material specific it contains ingredients which are important in the context of the superconducting transition metal oxide interfaces. Therefore, our study has some interesting implications related with the mentioned systems. Namely, in the low electron concentration regime which is realized in those systems the $extended$ $s$-$wave$ symmetry of the gap seems to be the most probable scenario at least within the real-space pairing approach. Such pairing symmetry leads to the appearance of the dome-like behavior of $T_c$ in a straightforward manner without the inclusion of electron-electron interaction effects and/or complex multiband phenomena. In such situation the stability of the non-zero momentum pairing in the form of the so-called helical state points to the transition metal oxide interfaces (e.g. LaAlO$_3$/SrTiO$_3$ or LaAlO$_3$/KTaO$_3$) as a promising candidates for the realization of the intrinsic supercurrent diode effect. Such non-reciprocal superconducting feature may be useful when it comes to applications in the field of modern superconducting electronics. Furthermore, it can also provide an experimental evidence for the appearance of the helical state itself. Namely, the fact that the critical currents are different in the opposite directions of a superconducting LAO/STO-based device can be considered as an indicator of the appearance of the non-zero Cooper pair momentum pairing. Since critical currents in the direction along and against the $\mathbf{Q}$ vector can be different only if $\mathbf{Q}\neq 0$. This concepts requires further detailed analysis which is beyond the scope of this work. We should see progress along this line soon.

\section{Acknowledgement}
This work was supported by National Science Centre, Poland (NCN) according to decision 2017/26/D/ST3/00109.

\appendix
\section{On-site real space pairing}
For the sake of completeness we have supplemented our results for the case of inter-site pairing scenario (provided in Section III ) with those for which the superconducting state results from an onsite pairing mechanism. In such case the pairing term has the following form
\begin{equation}
\begin{split}
 \hat{H}^U_{SC}&=-U\sum_{i}\hat{n}_{i\uparrow}\hat{n}_{i\downarrow},
 \end{split}
 \label{eq:Hamiltonian_pairing_onsite}
 \end{equation}
where $U>0$ corresponds to the coupling energy leading to the Cooper pairing in a straightforward manner. 

After the application of the Hartree-Fock-BCS approximation and the transformation to the reciprocal space one obtains the following form of term $\hat{H}^U_{SC}$
\begin{equation}
\begin{split}
 \hat{H}^U_{SC}&=\sum_{i}\big(\Delta\hat{c}^{\dagger}_{\mathbf{k}\uparrow}\hat{c}^{\dagger}_{-\mathbf{k}\downarrow}+\Delta^{\star}\hat{c}_{-\mathbf{k}\downarrow}\hat{c}_{\mathbf{k}\uparrow}\big)+\frac{N}{U}|\Delta|^2,
 \end{split}
 \label{eq:Hamiltonian_pairing_onsite_HFBCS}
 \end{equation}
where $N$ is the number of atomic sites and
\begin{equation}
    \Delta=-\frac{U}{N}\sum_k \hat{c}^{\dagger}_{\mathbf{k}\uparrow}\hat{c}^{\dagger}_{-\mathbf{k}\downarrow}.
\end{equation}
The main difference with regard to the situation analyzed in the main text is that due to the onsite character of the pairing mechanism the superconducting gap does not depend on $\mathbf{k}$ what corresponds to the so-called $s$-$wave$ pairing symmetry. The value of the gap calculated with the use of the Hartree-Fock-BCS approach for the case of Hamiltonian $\hat{H}=\hat{H}_0+\hat{H}^U_{SC}$ is presented in Fig. \ref{fig:mu_J_phase_diags} (c) and (f) as a function of both $U$ and $\mu$.

\section{Pairing in the helical bands}
Here we show how to express the gaps in helical bands in terms of the original spin-resolved gap amplitudes. In order to do that one first has to carry out the diagonalization procedure of the single particle Hamiltonian given by Eq. (\ref{eq:Hamiltonian_0}). The corresponding diagonalization transformation can be expressed in the following manner
\begin{equation}
\left(\begin{array}{c}
 \hat{c}_{\mathbf{k}\uparrow}  \\
\hat{c}_{\mathbf{k}\downarrow} \\
\end{array} \right)=\frac{1}{\sqrt{2}}
\left(\begin{array}{cc}
 -G_{\mathbf{k}} & G_{\mathbf{k}} \\
1 & 1 \\
\end{array} \right)
\left(\begin{array}{c}
 \hat{\alpha}_{\mathbf{k}-}  \\
\hat{\alpha}_{\mathbf{k}+} \\
\end{array} \right),
\label{eq:diag_transf}
\end{equation}
where 
\begin{equation}
    G_{\mathbf{k}}=\frac{1}{|\mathbf{\tilde{g}(k)}|}\Big(\tilde{g}_x(\mathbf{k})-i \tilde{g}_y(\mathbf{k})\Big),
    \label{eq:G_factor_appendix}
\end{equation}
while $\hat{\alpha}_{\mathbf{k}-}$ and $\hat{\alpha}_{\mathbf{k}+}$ are the annihilation operators in the two helical bands corresponding to the dispersion relations given by Eq. (\ref{eq:helical_bands}) and $\tilde{g}_{x/y}(\mathbf{k})$ are defined by Eq. (\ref{eq:g_tilda}). By applying the above transformation to the pairing term given by Eq. (\ref{eq:Hamiltonian_pairing_rec_space}) one arrives at 
\begin{equation}
\begin{split}
    \hat{H}_{SC}&=\frac{1}{2}\sum_{kl}\Big(\tilde{\Delta}^{(l)}_{\mathbf{k}\mathbf{Q}}\;\hat{\alpha}^{\dagger}_{\mathbf{k}l}\hat{\alpha}^{\dagger}_{-\mathbf{k+Q}l} + (\tilde{\Delta}^{(l)}_{\mathbf{k}\mathbf{Q}})^{\star}\;\hat{\alpha}_{-\mathbf{k+Q}l}\hat{\alpha}_{\mathbf{k}l}\Big)\\
    &+\frac{1}{J}\sum_{i<j}|\Delta^{\bar{\sigma}\sigma}_{ji}|,
    \label{eq:Hamiltonian_pairing_rec_helical_appendix}
\end{split}
\end{equation}
where the gaps in the helical bands are of the form
\begin{equation}
    \tilde{\Delta}^{(\pm)}_{\mathbf{k}\mathbf{Q}}=\pm\Big(G_{\mathbf{k}}^{\star}\Delta^{\uparrow\downarrow}_{\mathbf{k}\mathbf{Q}}+G_{\mathbf{-k+Q}}^{\star}\Delta^{\downarrow\uparrow}_{\mathbf{k}\mathbf{Q}}\Big).
    \label{eq:delta_helical_appendix}
\end{equation}
For simplicity in the above derivation we have assumed that only intra-helical-band pairing appears. This is justified by the fact that for relatively significant spin-orbit coupling energies the Fermi surfaces corresponding to the two helical bands are separated suppressing any inter-band pairing contributions. 

\section{Symmetry resolved gap amplitudes}
Here we show how the symmetry resolved real-space gap amplitudes defined by Eq. (\ref{eq:symmetry_factors}) are related with the $\mathbf{k}$-dependence of the resulting superconducting gap. As shown in the main text, the SC gap in reciprocal space depends on the real space gap amplitudes in the following manner
\begin{equation}
\begin{split}
    \Delta^{\bar{\sigma}\sigma}_{\mathbf{k}\mathbf{Q}}&=\sum_{i(j)}e^{i\mathbf{k}(\mathbf{R}_i-\mathbf{R}_j)}\tilde{\Delta}^{\bar{\sigma}\sigma}_{ji\mathbf{Q}}\\
    &=\tilde{\Delta}^{\bar{\sigma}\sigma}_{1,0}\;e^{ik_x}+\tilde{\Delta}^{\bar{\sigma}\sigma}_{-1,0}\;e^{-ik_x}\\
    &+\tilde{\Delta}^{\bar{\sigma}\sigma}_{0,1}\;e^{ik_y}+\tilde{\Delta}^{\bar{\sigma}\sigma}_{0,-1}\;e^{-ik_y}.
    \label{eq:delta_kQ_Appendix_C}
\end{split}
\end{equation}
In the following, we introduce the real-space symmetry resolved gap amplitudes
\begin{equation}
    \begin{split}
    \Delta^{\bar{\sigma}\sigma}_s &=             (\Delta^{\bar{\sigma}\sigma}_{1,0}+\Delta^{\bar{\sigma}\sigma}_{0,1}+\Delta^{\bar{\sigma}\sigma}_{-1,0}    +\Delta^{\bar{\sigma}\sigma}_{0,-1})/8,\\
        \Delta^{\bar{\sigma}\sigma}_d &= (\Delta^{\bar{\sigma}\sigma}_{1,0}-\Delta^{\bar{\sigma}\sigma}_{0,1}+\Delta^{\bar{\sigma}\sigma}_{-1,0}-\Delta^{\bar{\sigma}\sigma}_{0,-1})/8,\\
        \Delta^{\bar{\sigma}\sigma}_{p_x} &= (\Delta^{\bar{\sigma}\sigma}_{1,0}-\Delta^{\bar{\sigma}\sigma}_{-1,0})/4,\\
        \Delta^{\bar{\sigma}\sigma}_{p_y} &= (\Delta^{\bar{\sigma}\sigma}_{0,1}-\Delta^{\bar{\sigma}\sigma}_{0,-1})/4,
    \end{split}
    \label{eq:gap_symmetries_k_space}
\end{equation}
which correspond to the $extended$ $s$-$wave$, $d$-$wave$, $p_x$-$wave$, and $p_y$-$wave$ pairing symmetries, respectively. It should be noted that in general the symmetry of the gap does not have to be identical with the symmetry of the lattice or the symmetry of the orbitals which are use to construct the single particle part of the model Hamiltonian. 

The superconducting gap in reciprocal space given by Eq. (\ref{eq:delta_kQ_Appendix_C}) can by expressed in terms of the symmetry resolved gap parameters as follows
\begin{equation}
\begin{split}
    \Delta^{\bar{\sigma}\sigma}_{\mathbf{k}\mathbf{Q}}&=\Delta^{\bar{\sigma}\sigma}_s\;\big(\cos(k_x)+\cos(k_y)\big)\\
    &+\Delta^{\bar{\sigma}\sigma}_d\;\big(\cos(k_x)-\cos(k_y)\big)\\
    &+i\Delta^{\bar{\sigma}\sigma}_{p_x}\;\sin(k_x)\\
    &+i\Delta^{\bar{\sigma}\sigma}_{p_y}\;\sin(k_y).\\
    \label{eq:delta_kQ_symmtery_factors}
\end{split}
\end{equation}
From the above one can see that in general we can obtain a mixture of gap symmetries. 
By substituting from the above equations to the the expression for the gap amplitudes in the helical bands given by Eq. (\ref{eq:delta_helical}) [or Eq. (\ref{eq:delta_helical_zero_mementum}) for the $\mathbf{B}=0$ and $\mathbf{Q=0}$ case] one can determine how the real-space gap amplitudes determine the $\mathbf{k}$-space ($extended$ $s$-$wave$, $d$-$wave$, $p_x$/$p_y$-$wave$) dependence of the resulting superconducting gap. 

It should be noted that the spin degree of freedom also plays a role when it comes to the resulting symmetry of the gap since the superconducting order parameter needs to be antisymmetric with respect to the interchange of particles as our system is composed of fermions. Therefore, the spin-singlet state (odd) of the Cooper pairs can be realized in connection with the $extended$ $s$-$wave$ or $d$-$wave$ symmetries (even) and the spin-triplet state (even) of the Cooper pair can be realized with the $p_x$/$p_y$-$wave$ symmetry (odd). In order to take this into account we have introduced the symmetry resolved gap amplitudes which take into account both spin and real-space degrees of freedom and are defined in the following manner
\begin{equation}
    \begin{split}
        \Delta^s_s &= \Delta^{\uparrow\downarrow}_s - \Delta^{\downarrow\uparrow}_s,\\
        \Delta^s_d &= \Delta^{\uparrow\downarrow}_d - \Delta^{\downarrow\uparrow}_d,\\
        \Delta^t_{p_x} &= \Delta^{\uparrow\downarrow}_{p_x} + \Delta^{\downarrow\uparrow}_{p_x},\\
        \Delta^t_{p_y} &= \Delta^{\uparrow\downarrow}_{p_y} + \Delta^{\downarrow\uparrow}_{p_y},\\
    \end{split}
\end{equation}
which are in fact the same as Eqs. (\ref{eq:gap_symmetries}), but here we use the notation provided by Eqs. (\ref{eq:gap_symmetries_k_space}). In the case when both $\mathbf{B}=0$ and $\alpha=0$, only spin-singlet pairing appears and only $extended$ $s$-$wave$ or $d$-$wave$ symmetries are possible. However, for the case of nonzero external magnetic field and spin-orbit coupling a singlet-triplet mixing appears and several contribution to the gap may be realized which is reflected by Eq. (\ref{eq:delta_kQ_symmtery_factors}).

\newpage

\bibliography{refs.bib}

\end{document}